\documentclass[12pt,fleqn]{article}
\usepackage{color,amsmath,epsfig}

\newcommand{\be}{\begin{eqnarray}}
\newcommand{\ee}{\end{eqnarray}}
\newcommand{\nee}{\nonumber\end{eqnarray}}

\newcommand{\noi}{\noindent}

\newcommand{\mbf}      {\boldmath}

\newcommand{\eq}[1]  {\mbox{(\ref{eq:#1})}}
\newcommand{\fig}[1] {\mbox{Fig.~\ref{fig:#1}}}
\newcommand{\Fig}[1] {\mbox{Figure~\ref{fig:#1}}}

\def\b               {\beta}

\def\x               {\chi}
\def\D               {\Delta}

\def\ti              {\tilde}

\def\sq              {\ti q}
\def\st              {\ti t}
\def\sb              {\ti b}
\def\ch              {\ti \x^\pm}
\def\chp             {\ti \x^+}

\def\nt              {\ti \x^0}
\def\sg              {\ti g}

\newcommand{\mst}[1]   {m_{\st_{#1}}}

\def\DR              {${\rm\overline{DR}}$}

\newcommand{\gsim}{\;\raisebox{-0.9ex}
           {$\textstyle\stackrel{\textstyle >}{\sim}$}\;}
\newcommand{\lsim}{\;\raisebox{-0.9ex}{$\textstyle\stackrel{\textstyle<}
           {\sim}$}\;}

\definecolor{red}{rgb}{1,0,0}
\definecolor{lred}{rgb}{1,0.15,0}
\definecolor{dred}{rgb}{0.7,0,0}
\definecolor{ddred}{rgb}{0.5,0,0}

\definecolor{green}{rgb}{0,1,0}
\definecolor{lgreen}{rgb}{0.3,1,0}
\definecolor{dgreen}{rgb}{0,0.7,0}

\definecolor{blue}{rgb}{0,0,1}
\definecolor{lblue}{rgb}{0,0.5,1}
\definecolor{llblue}{rgb}{0,0.8,1}
\definecolor{dblue}{rgb}{0,0,0.7}
\definecolor{ddblue}{rgb}{0,0,0.5}

\definecolor{orange}{rgb}{1,0.5,0}
\definecolor{dorange}{rgb}{0.8,0.3,0}

\definecolor{pink}{rgb}{1,0,1}
\definecolor{lila}{rgb}{0.7,0,0.7}
\definecolor{violet}{rgb}{0.7,0,0.7}

\definecolor{gray}{rgb}{0.2,0.2,0.2}

\begin{document}


\begin{center}

{\large\bf\boldmath
    Comparison of SUSY mass spectrum calculations}\\[6mm]

B. Allanach, \underline{S. Kraml},\footnote{Speaker} W. Porod\\[6mm]

\end{center}

\begin{abstract}
  We provide a comparison of the results of four SUSY 
  mass spectrum calculations in mSUGRA: Isajet, SuSpect, 
  SoftSusy, and SPheno. In particular, we focus on the 
  high $\tan\b$ and focus point regions, where the differences 
  in the results are known to be large. 
\end{abstract}

\begin{center}\noi
{\small \it Contribution to SUSY02, 
10th International Conference on Supersymmetry and Unification\\ 
of Fundamental Interactions, 17--23 June 2002, DESY Hamburg, Germany.}
\end{center}

\section{Introduction}

Many SUSY studies rely on computer codes that calculate the  
mass spectrum of the minimal supersymmetric standard model (MSSM),
the couplings, branching ratios, {\it etc.}, from given sets of model 
parameters. For the LHC, for instance, many simulations are done for 
particular benchmark scenarios or by mapping the $(m_0,m_{1/2})$ 
parameter plane. For such studies it is certainly important wheather 
a particular decay channel is open or not and what branching ratio 
it has. Also, theoretically or experimentally excluded regions depend 
on the details of the spectrum. 
Studies for an $e^+e^-$ Linear Collider deal, in addition, with high 
precision measurements of (s)particle properties, with the 
determination of the underlying SUSY breaking parameters, their 
extrapolation to the GUT scale, model destinction, {\it etc}. 
Experimental accuracies of the per-cent or even per-mille level 
are expected. 
It is thus clear that we need theoretical predictions 
of a precision comparable to the experimental accuracy.
However, it has been noticed \cite{Allanach:2001hm,Ghodbane:2002kg} 
that different programs can give quite different results for the same 
set of input parameters. 

In this article, we compare the mass spectrum calculations 
of four public codes: Isajet\,7.63~\cite{Baer:1999sp}, 
SuSpect\,2.005~\cite{suspect}, SoftSusy\,1.4~\cite{Allanach:2001kg}, 
and SPheno\,1.0~\cite{spheno}, 
in the minimal supergravity (mSUGRA) framework. 
We discuss the renormalization group (RG) running and the 
implementation of radiative corrections, 
concentrating on the parameter regions where the largest 
differences are encountered: large $\tan\b$ and large $m_0$.
An overview of which corrections are implemented in each of 
the four programs is given in Table~1.

\begin{table}
\begin{center}
\footnotesize{\begin{tabular}{|c|c|c|c|c|}
  \hline
    & \bf Isajet\,7.63 & \bf SuSpect\,2.005 
    & \bf SoftSusy\,1.4 & \bf SPheno\,1.0 \\
  \hline
  \bf RGEs & & & & \\
    gauge + Yuk. & 2--loop & 2--loop & 2--loop & 2--loop  \\ 
    gaugino par. & 2--loop & 2--loop & 2--loop & 2--loop  \\ 
    scalar par.  & 2--loop & 1--loop & 1--loop & 2--loop  \\ 
  \hline
  \bf SUSY\ masses & & & & \\ 
    $\ch,\nt$ & some corr. for $\ch_1$ 
      & \multicolumn{2}{c|}{1--loop approx. for $\D M_1$, $\D M_2$, $\D\mu$} 
      & full 1--loop \\
    $\st$ & --- & $\st g + t\sg$ + Yuk. & full 1--loop & full 1--loop \\
    $\sb$ & --- & $\sb g + b\sg$  & full 1--loop & full 1--loop \\
    $\sg$ & \multicolumn{4}{c|}{$g\sg + q\sq$ loops resummed} \\
  \hline
  \bf Yukawa cpl.& & & & \\
  $h_t$ & full 1--loop resum.         
                  & $tg+\st\sg$ & full 1--loop & full 1--loop \\
  $h_b$ & full 1--loop resum.         
                  & \multicolumn{2}{c|}{$bg + \sb\sg + \st\ch$ corr. resummed} 
                  & full 1--loop resum. \\
  \hline
  \bf Higgs sector & & \multicolumn{3}{c|}{} \\
    tadpoles & 3rd gen.\ (s)fermions 
      & \multicolumn{3}{c|}{complete 1--loop corrections \cite{Pierce:1996zz}} \\
    $h^0$, $H^0$ & 1--loop  \cite{Bisset:1995dc} 
                 & 1--loop  \cite{Carena:1995wu}
                 & 2--loop  \cite{Heinemeyer:1998yj}
                 & 2--loop  \cite{Brignole:2001jy} \\
  \hline 
\end{tabular}}
\end{center}
\caption{RGEs and radiative corrections implemented in Isajet, SuSpect, 
SoftSusy, and SPheno.}
\end{table}

\section{Large \mbf $\tan\b$}

Large $\tan\b$ has always been recognized as a difficult case since  
it requires a thorough treatement of the bottom Yukawa coupling $h_b$.  
It is well known \cite{Hempfling:1993kv} 
that $h_b$ gets large $\tan\b$ enhanced corrections 
from SUSY loops, the dominant contributions coming from $\sb\sg$ and 
$\st\chp$ exchanges. 
These generate a $H^0_2 b\bar b$ coupling, which is forbidden at tree-level, 
${\cal L}\sim h_b\,H^0_1 b\bar b + \Delta h_b\, H^0_2 b\bar b$.   
This modifies the tree-level relation between the bottom mass and 
Yukawa coupling, $m_b=h_bv_1\to m_b=h_bv_1(1+\Delta_b)$ with 
$\Delta_b = (\Delta h_b/h_b)\tan\b$. 
In the programs under discussion this is taken into account as 
\begin{equation}
  h_b(M_Z)=\hat m_b^{\rm MSSM}(M_Z)/v_1(M_Z) \,, \quad 
  \hat m_b^{\rm MSSM}(M_Z) = 
  \frac{\hat m_b^{\rm SM}(M_Z)}{1 + \D m_b/m_b}\,.
\label{eq:mbMSSM}
\end{equation} 
Here $\hat m_b^{\rm SM}$ is the 
\DR\ bottom mass in the Standard Model  
and $\D m_b = (\D m_b)^{\,\sb\sg+\st\chp+...}$ 
contains the SUSY-loop corrections. 
The complete 1-loop expression for $\D m_b$ is given 
in \cite{Pierce:1996zz}.\footnote{Here note that 
  \cite{Baer:1999sp,suspect,Allanach:2001kg,spheno} and 
  \cite{Pierce:1996zz} partly have different conventions, 
  {\it e.g.}, for the ordering of the squark mass 
  eigenstates and the sign of $\mu$.}
Compared to the naive 1-loop expansion 
$\hat m_b^{\rm MSSM} = \hat m_b^{\rm SM}(1-\D m_b/m_b)$, 
eq.~\eq{mbMSSM} makes a numerical difference of about $10\%$ in $h_b$ 
and about 10--30\% in $m_A$ for large $\tan\b$. 
The re-summation of SUSY threshold corrections \cite{Carena:1999py} 
will be discussed elsewhere~\cite{inprep}.
Although all four programs now apply eq.~\eq{mbMSSM}, 
some numerical differences in $h_b$ remain.
These are partly due to differences in $\alpha_s$: 
Suspect, SoftSusy and SPheno calculate $\alpha_s$ in the 
\DR\ scheme, Isajet uses the $\overline{\rm MS}$ value. 
Another reason is that Isajet uses $m_b=m_b(M_{SUSY})$ for  
the expression $\Delta m_b/m_b$ in eq.~\eq{mbMSSM}, 
while the other programs use $m_b(M_Z)$ or the bottom pole mass; 
also the gluino masses differ by about 5\%. 
Moreover, the vacuum expectation 
values $v_{1,2}$ are not running in Isajet. 

\begin{figure}[ht!]
\begin{center} {\setlength{\unitlength}{1mm}
\begin{picture}(128,64)
\put(0,0){\mbox{\epsfig{figure=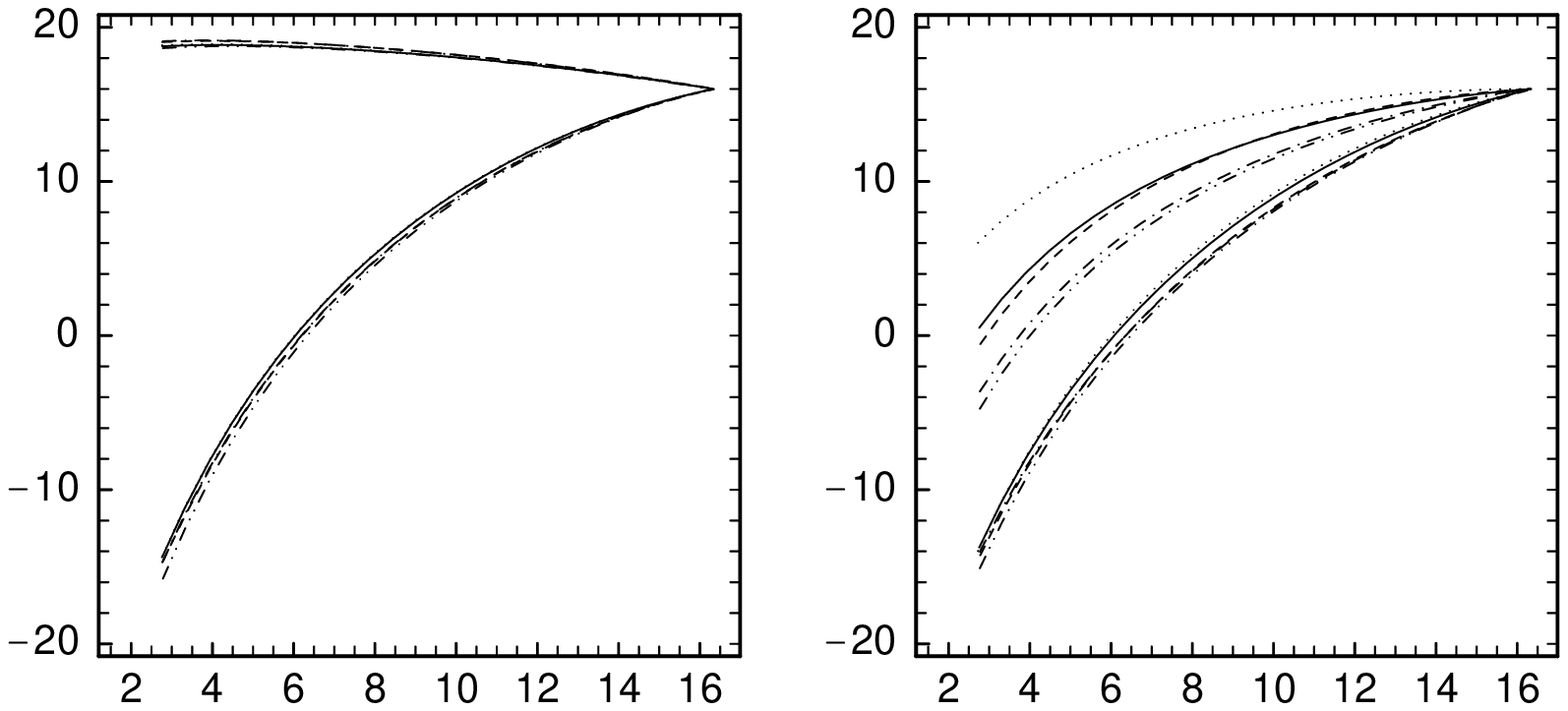,height=6cm}}}
\put(5,60){\mbox{$m_{H_i}^2~[10^4~{\rm GeV}^2]$}}
\put(68,60){\mbox{$m_{H_i}^2~[10^4~{\rm GeV}^2]$}}
\put(22,-2){\mbox{Log($Q$/1~GeV)}}
\put(85,-2){\mbox{Log($Q$/1~GeV)}}
\put(20,49){\mbox{$m_{H_1}^2$}}
\put(81,49){\mbox{$m_{H_1}^2$}}
\put(28,26){\mbox{$m_{H_2}^2$}}
\put(91,26){\mbox{$m_{H_2}^2$}}
\put(35,12){\mbox{\mbf $\tan\b=10$}}
\put(97,12){\mbox{\mbf $\tan\b=50$}}
\end{picture} }
\end{center}
\caption{Running of $m_{H_{1,2}}^2$ as a function of the scale $Q$, for 
  $m_0=400$~GeV, $m_{1/2}=300$~GeV, $A_0=0$, $\mu>0$, $\tan\b=\{10,50\}$,
  $M_t=175$~GeV. The full (dotted) lines are for Isajet\,7.63 (7.58), 
  the dashed lines are for SuSpect\,2.005, the dash-dotted ones for 
  SoftSusy\,1.4, and the dash-dot-dotted ones for SPheno\,1.0.
\label{fig:runMH2SPS4}}
\end{figure}

The bottom Yukawa coupling has its largest 
effect in the Higgs sector. 
\Fig{runMH2SPS4} shows the running of $m_{H_{1,2}}^2$ for $m_0=400$~GeV, 
$m_{1/2}=300$~GeV, $A_0=0$, $\mu>0$, and the two cases $\tan\b=10$ 
and $\tan\b=50$. As one can see, there is good agreement for not too 
large $\tan\b$. However, for $\tan\b=50$, quite different results are 
obtained for $m_{H_1}^2$, whose evolution is driven by $h_b$:
\begin{equation}
  \frac{dm_{H_1}^2}{dt} \sim \frac{3}{8\pi^2}\,h_b\,X_b +\ldots\,,\quad  
  X_b = (m_{\ti Q}^2 + m_{\ti D}^2 + m_{H_1}^2 + A_b^2)\,.
\end{equation}
Note in particular the dotted line which shows the result
obtained with Isajet\,7.58. In this version, the SUSY corrections 
to $h_b$ were not yet resummed.

The differences in $m_{H_1}^2$ directly translate into $m_A^2$ 
and thus into the physical Higgs boson masses, since
\begin{equation}
   m_A^2 = \frac{1}{c_{2\b}}
   \left( \overline{m}_{H_2}^2 - \overline{m}_{H_1}^2\right) 
   + \frac{s_\b^2\,t_1}{v_1} + \frac{c_\b^2\,t_2}{v_2} - M_Z^2  \,.
\label{eq:mA}
\end{equation}
Here $\overline{m}_{H_i}^2 = m_{H_i}^2 - t_i/v_i$, $i=1,2$, and 
$t_{1,2}$ are the tadpole contributions. The self energies of 
$Z$ and $A$ have been neglected in eq.~\eq{mA}. 
We note that including only the tadpoles from the third generation 
is in general a good approximation. The remaining 1-loop contributions 
account for a ${\cal O}(1\%)$ correction. 

\begin{figure}[ht!]
\begin{center} {\setlength{\unitlength}{1mm}
\begin{picture}(145,135)
\put(4,0){\mbox{\epsfig{figure=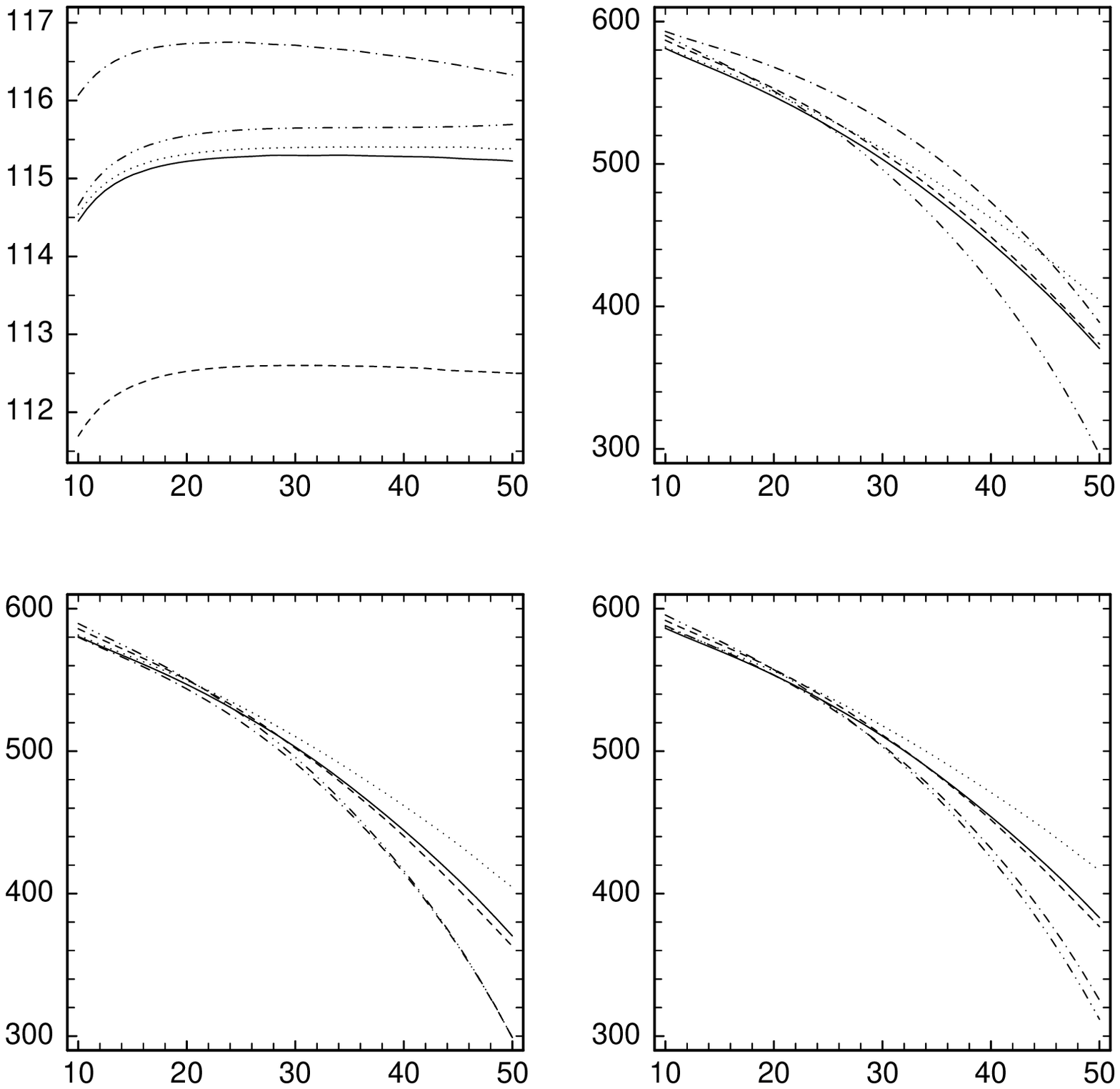,width=14cm}}}
\put(1,100){\rotatebox{90}{$m$~[GeV]}}
\put(72,100){\rotatebox{90}{$m$~[GeV]}}
\put(1,30){\rotatebox{90}{$m$~[GeV]}}
\put(72,30){\rotatebox{90}{$m$~[GeV]}}
\put(38,1){\mbox{$\tan\b$}}
\put(108,1){\mbox{$\tan\b$}}
\put(38,71){\mbox{$\tan\b$}}
\put(108,71){\mbox{$\tan\b$}}
\put(61,84){\mbox{\mbf $h^0$}}
\put(128,126){\mbox{\mbf $H^0$}}
\put(60,56){\mbox{\mbf $A^0$}}
\put(128,56){\mbox{\mbf $H^\pm$}}
\end{picture} }
\end{center}
\caption{Higgs boson masses as a function of $\tan\b$, for 
  $m_0=400$~Gev, $m_{1/2}=300$~GeV, $A_0=0$, $\mu>0$, $M_t=175$~GeV; 
  full (dotted) lines: Isajet\,7.63 (7.58), dashed: SuSpect\,2.005, 
  dash-dotted: SoftSusy\,1.4, dash-dot-dotted: SPheno\,1.0.
\label{fig:higgsmasses}}
\end{figure}

\Fig{higgsmasses} shows the Higgs boson masses obtained by the four 
programs as a function of $\tan\b$. 
The new Isajet version 7.63 
has led to a majer improvement compared to the situation 
discussed in \cite{Ghodbane:2002kg,stmalo} (the results obtained by 
Isajet\,7.58 are again shown as dotted lines in \fig{higgsmasses}). 
For $m_A$ and $m_{H^\pm}$ there is now agreement within 
$\sim10\%$ up to $\tan\b\sim45$. 
Sources for the remaining differences are pointed out above. 
Moreover, it makes a difference whether one uses running 
couplings and/or masses for the tadpoles $t_{1,2}$. 
Here each program has a different approach.
For the neutral scalars, however, the situation is not so good. 
Especially for $m_{h^0}$, a discrepancy of $\sim 4$~GeV is too large 
compared to the expected experimental accuracy. This discrepancy 
is mainly due to the different radiative corrections taken into 
account for the ($h^0,H^0$) system. 
They vary between 1- and 2-loop, effective potential 
and diagramatic calculations, see Table~1. 
Given the expected experimental accuracy for $m_{h^0}$ it is clear 
that the best available calculation should be used.


\section{Large \mbf $m_0$}

\begin{figure}[t!]
\begin{center} {\setlength{\unitlength}{1mm}
\begin{picture}(60,60)
\put(0,0){\mbox{\epsfig{figure=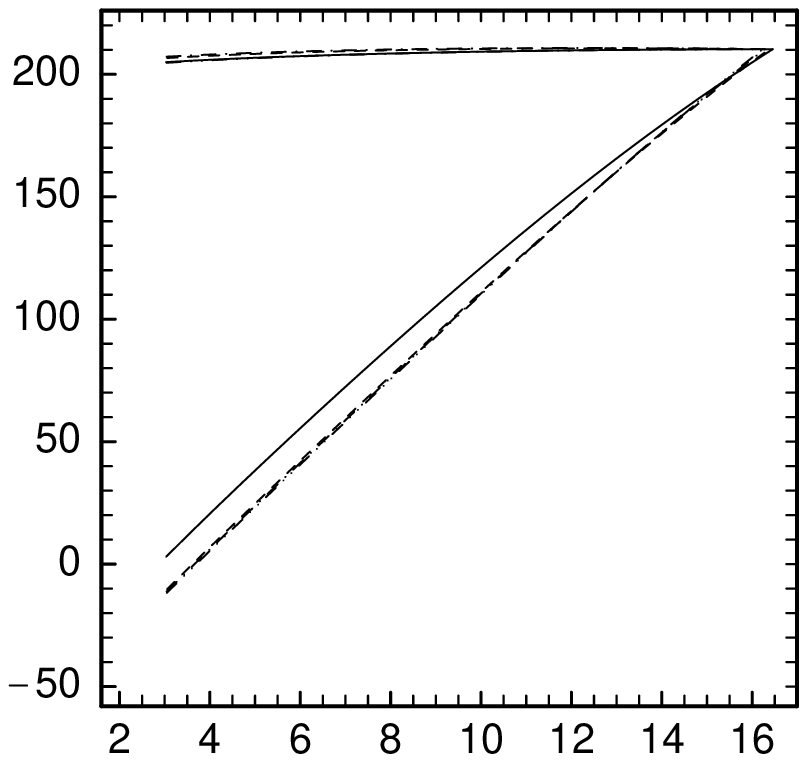,height=6cm}}}
\put(-6,19){\rotatebox{90}{$m_{H_i}^2~[10^4~{\rm GeV}^2]$}}
\put(20.5,-3){\mbox{Log($Q$/1~GeV)}}
\put(20,49){\mbox{$m_{H_1}^2$}}
\put(32,26){\mbox{$m_{H_2}^2$}}
\put(-6,56){\mbox{\bf a)}}
\end{picture} 
\hspace{10mm}
\begin{picture}(60,60)
\put(0,-1){\mbox{\epsfig{figure=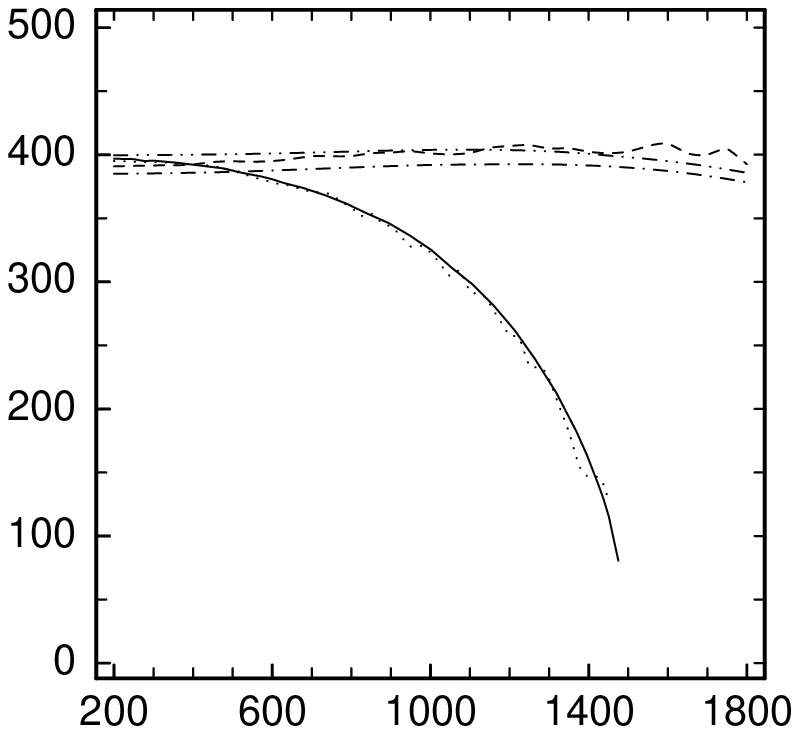,height=6.2cm}}}
\put(-7,27){\rotatebox{90}{$\mu$~[GeV]}}
\put(25,-3){\mbox{$m_0$~[GeV]}}
\put(-6,56){\mbox{\bf b)}}
\end{picture} }
\end{center}
\caption{{\bf a)} Running of $m_{H_{1,2}}^2$ for $m_0=1450$~GeV; 
  {\bf b)} $\mu$ as a function of $m_0$; 
  for $m_{1/2}=300$~GeV, $A_0=0$, $\mu>0$, 
  $\tan\b=10$, and $M_t=175$~GeV; 
  full (dotted) lines: Isajet\,7.63 (7.58), dashed: SuSpect\,2.005, 
  dash-dotted: SoftSusy\,1.4, dash-dot-dotted: SPheno\,1.0.
\label{fig:SPS2}}
\end{figure}

For large $m_0$, the running of $m_{H_2}^2$ becomes very steep and 
very sensitive to the top Yukawa coupling $h_t=\hat m_t/v_2$\,:
\begin{equation}
  \frac{dm_{H_2}^2}{dt} \sim \frac{3}{8\pi^2}\,h_t\,X_t +\ldots\,,\quad  
  X_t = (m_{\ti Q}^2 + m_{\ti U}^2 + m_{H_2}^2 + A_t^2)\,.
\end{equation}
As a result, the $\mu$ parameter given by 
\begin{equation}
   \mu^2 = \frac{\overline{m}_{H_1}-\overline{m}_{H_2}^2\tan^2\b}
                {\tan^2\b-1} - \frac{1}{2}\,M_Z^2  
\label{eq:muex}
\end{equation}
becomes extremely sensitive to $h_t$. 
This is visualized in \fig{SPS2} where we show in   
(a) the running of $m_{H_{1,2}}^2$ for $m_0=1450$~GeV, and in 
(b) $\mu$ as a function of $m_0$. 
The other parameters are $m_{1/2}=300$~GeV, $A_0=0$, $\mu>0$, 
and $\tan\b=10$. The large discrepancy in $\mu$ for $m_0\gsim 800$~GeV 
lead to completely different chargino/neutralino properties and 
likewise to very different excluded regions 
in Isajet compared to the other programs. 
For instance, radiative EWSB breaks down in Isajet 
for $m_0\gsim 1.5$~TeV. 
In SuSpect, SoftSusy, and SPheno, this happens only for 
$m_0\gsim 2.5$--2.8~TeV.\footnote{After the conference, a sign error 
was corrected in SPheno. As a consequence, its results for 
large $m_0$ now nicely agree with those of SoftSusy and SuSpect 
(contrary to what was presented in the talk).}

\begin{figure}[ht!]
\begin{center} {\setlength{\unitlength}{1mm}
\begin{picture}(61,66)
\put(0,0){\mbox{\epsfig{figure=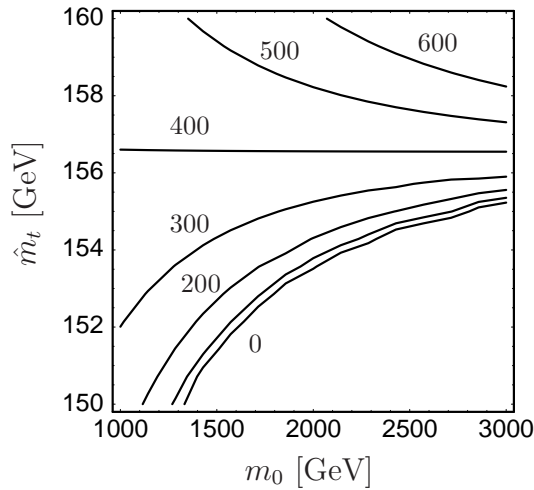,height=6.4cm}}}
\put(-6,26.5){\rotatebox{90}{$\hat m_t$~[GeV]}}
\put(25,-2){\mbox{$m_0$~[GeV]}}
\put(25.5,15){\mbox{\footnotesize 0}}
\put(16.5,23){\mbox{\footnotesize 200}}
\put(15,31){\mbox{\footnotesize 300}}
\put(15,44){\mbox{\footnotesize 400}}
\put(27,54){\mbox{\footnotesize 500}}
\put(48,55){\mbox{\footnotesize 600}}
\end{picture} }
\end{center}
\caption{The parameter $\mu$ as given by eq.~\eq{muap} in the 
  ($m_0,\hat m_t$) plane, for $m_{1/2}=300$~GeV, $A_0=0$, 
   and $\tan\b=10$.
\label{fig:mu_m0mt}}
\end{figure}

In order to understand the behaviour in \fig{SPS2}b it is useful 
to write eq.~\eq{muex} in the form 
\begin{equation}
   \mu^2 \simeq c_1\, m_0^2 + c_2\, m_{1/2}^2 - 0.5 M_Z^2\,.
\label{eq:muap}
\end{equation}
Approximate analytical expressions for $c_1$ and $c_2$ can be found 
{\it e.g.}, in \cite{Carena:1994bv, Drees:1995hj}. 
For $A_0=0$ and $\tan\b=10$ we get \cite{Drees:1995hj}
\begin{equation}
   c_1 \sim\, \left(\frac{\hat m_t}{156.5~{\rm GeV}}\right)^2-1\,,\qquad 
   c_2 \sim\, \left(\frac{\hat m_t}{102.5~{\rm GeV}}\right)^2-0.52\,.
\label{eq:c1c2}
\end{equation}
Since the Higgs potential is minimized at 
$M_{SUSY}=\sqrt{\mst{1}\mst{2}}$, we take $\hat m_t$ in eq.~\eq{c1c2} 
as $\hat m_t=\hat m_t(M_{SUSY})$. The $m_0$ dependence seen in Isajet 
is reproduced for $\hat m_t\sim 151$~GeV. The one of SuSpect, 
SoftSusy and SPheno is reproduced for $\hat m_t\sim 155$~GeV. 
\Fig{mu_m0mt} shows a contour plot of $\mu$ in the ($m_0,\hat m_t$) plane. 
Notice the fast increasing dependence on $\hat m_t$ for increasing $m_0$. 
Notice also that for $\hat m_t\sim 156$--157~GeV, $\mu$ becomes almost 
independent of $m_0$, which is the actual focus point condition. 

There are some obvious differences in the calculations. 
For instance, $M_{SUSY}$, the scale where the SUSY parameters 
are frozen out and the Higgs potential is minimized, varies 
by about 100~GeV due to different radiative corrections 
to the stop masses, {\it c.f.} Table~1.
In the loop corrections to $m_t$, analogous differences occur 
as discussed above for $\Delta m_b/m_b$.
Also the evolution of $h_t$ between $M_Z$ and $M_t$ and the inclusion 
of threshold effects are delicate points. 
However, this is not yet sufficient to explain the observed 
discrepancies. 
More work is needed to clarify the situation.

\section{Conclusions}

For the calculation of the SUSY mass spectrum from GUT scale boundary 
conditions, there are two particular difficult parameter regions 
where large numerical differences have been noticed: large $\tan\b$ 
and large $m_0$. These regions are very sensitive to the bottom 
and top Yukawa couplings, respectively. 

The inclusion of the SUSY 1-loop corrections to $h_b$ 
has led to a considerable improvement in the large $\tan\b$ case. 
In particular, the four programs now agree on $m_A$ within $\lsim 10\%$ 
for $\tan\b\lsim 45$. Further improvements are of course desirable.

For large $m_0$, there are still very large numerical discrepancies 
due to the corrections to $h_t$. 
As a matter of fact, $h_t$ is much smaller in Isajet than in the other 
programs. Some differences in the calculation of $h_t$ have been pointed 
out, but these do not satisfingly explain the observed discrepancies.
Work is in progress to clarify the situation~\cite{inprep}.

\section*{Acknowledgements}

We thank H.~Baer, A.~Djouadi, and J.-L.~Kneur for useful discussions.


\end{document}